\documentclass[a4paper]{article}
\usepackage[left=2.5cm, right=2.5cm, top=2.5cm, bottom=2.5cm]{geometry}

\newcommand{\changefontsize}{\fontsize{12}{16}\selectfont}

\changefontsize

\usepackage{cmbright}
\usepackage[T1]{fontenc}

\usepackage[utf8]{inputenc}
\usepackage{amssymb}
\usepackage{amsfonts}
\usepackage{amsmath}
\usepackage{fontsmpl}
\usepackage{graphicx}
\usepackage{epsfig}
\usepackage[utf8]{inputenc}
\usepackage[english]{babel}
 
\usepackage{amsthm}
\usepackage{subcaption}
\usepackage{epsfig}

\usepackage[table,xcdraw]{xcolor}  %need this for coloring rows in tables
 \usepackage{booktabs} % need for the tables
\usepackage{caption} 
\usepackage{subcaption}
\usepackage{rotating}

\usepackage[subfigure]{tocloft}

\usepackage{tikz}
\usepackage{subcaption}
\usepackage{xcolor}

\usepackage{booktabs}
\usepackage{multirow}
\usepackage{rotating}
\usepackage{forest}
\usepackage{xcolor}

\usepackage{graphicx}
\usepackage{booktabs}
\usepackage{float}

\usepackage{amsmath}

\title{Artificial General Intelligence and the End of Human Employment: The Need to Renegotiate the Social Contract}
\author{Pascal Stiefenhofer\\ Department of Economics, Newcastle University\\pascal.stiefenhofer@newcastle.ac.uk}
\date{January 2025}

\begin{document}

	\maketitle

\begin{abstract}
The emergence of Artificial General Intelligence (AGI) labor, including AI agents and autonomous systems operating at near-zero marginal cost, reduces the marginal productivity of human labor, ultimately pushing wages toward zero. As AGI labor and capital replace human workers, economic power shifts to capital owners, resulting in extreme wealth concentration, rising inequality, and reduced social mobility. The collapse of human wages causes aggregate demand to deteriorate, creating a paradox where firms produce more using AGI, yet fewer consumers can afford to buy goods. To prevent economic and social instability, new economic structures must emerge, such as Universal Basic Income (UBI), which redistributes AGI-generated wealth, public or cooperative AGI ownership, ensuring broader access to AI-driven profits, and progressive AGI capital taxation, which mitigates inequality and sustains aggregate demand. Addressing these challenges  in form of renegotiation the Social Contract\footnote{The Social Contract, originally titled On the Social Contract; or, Principles of Political Right (French: Du contrat social; ou, Principes du droit politique), is a 1762 work by Swiss philosopher Jean-Jacques Rousseau. In this seminal text, Rousseau examines the nature of political legitimacy and proposes a model in which authority is derived from the collective will of the people, ensuring both governance and individual freedom. The book addresses the tensions of commercial society, a theme Rousseau had previously explored in Discourse on Inequality (1755), and offers a vision for a just and equitable political order based on social cooperation and mutual obligation.} is crucial to maintaining economic stability in a post-labor economy.
\end{abstract}

\


\section{Introduction}

The rapid emergence of Artificial General Intelligence (AGI) marks a paradigm shift in production, labor dynamics, and economic power\cite{terrile2019}. Unlike past technological advancements, which primarily enhanced human productivity, AGI possesses the capability to fully replace both cognitive and physical labor \cite{feng2024}. This unprecedented shift threatens to render human employment obsolete across numerous industries. Operating at near-zero marginal cost and continuously improving through self-learning, AGI offers firms an overwhelmingly efficient alternative to human workers. As a result, labor demand is poised to collapse, triggering a downward spiral in wages and fundamentally disrupting the historical equilibrium between labor and capital. In this new economic order, ownership of AGI assets—rather than human labor—becomes the primary determinant of wealth and power.  

The implications of this transformation are profound. Should human wages approach zero, traditional mechanisms of wealth distribution and economic participation become unsustainable. The classical capitalist model, predicated on wage-based consumption, faces a paradox: firms achieve unprecedented productivity while consumer purchasing power erodes due to mass unemployment. This dynamic threatens to destabilize markets, deepen economic inequality, and create a stark divide between AGI capital owners and those excluded from economic participation.  

The collapse of wage-based employment presents an urgent question: how should society adapt to an economy where human labor is no longer the primary means of income distribution? This paper explores the economic ramifications of AGI-driven automation and the policy interventions necessary to prevent systemic collapse. To ensure that AGI’s productivity gains benefit society as a whole rather than an elite minority, the existing social contract must be reimagined\cite{rousseau1762}. Without proactive policy responses, unchecked AGI-driven automation risks economic stagnation, extreme wealth polarization, and widespread social disruption.  

A critical lens for analyzing these transformations is the Cobb-Douglas production function, a foundational model in production economics that captures the relationship between labor, capital, and output \cite{wicksteed1894}\cite{cobb1928}\cite{vale2024}. Characterized by its elasticities—represented as exponents—the function quantifies how changes in input levels affect output. While widely applied in both economic theory and empirical modeling \cite{vale2024}\cite{oglih2024}\cite{stiefenhofer2013,stiefenhofer2014,stiefenhofer2016,stiefenhofer2017}, the Cobb-Douglas framework has yet to be extended to account for AGI’s impact on labor demand. As AGI becomes deeply integrated into production processes, its ability to dynamically adjust these elasticities in real time could significantly enhance total factor productivity TFP, improving overall efficiency and addressing resource misallocation \cite{stiefenhofer2024}.  

This paper develops a novel production model that incorporates AGI-driven capital and labor. Section 2 introduces the foundational model integrating AGI capital, while Section 3 extends the framework to include AGI labor. Section 4 examines the combined effects of AGI labor and capital on production dynamics. Section 5 analysis the powershift towards AGI capital owners.  Finally, the paper concludes with a discussion on broader economic implications and the policy measures required to navigate this unprecedented shift in the labor economy.

\section{Model I: Production Model with Human Labor, Capital  and AGI as a form of Capital}
	
With the advancement of , the fundamental nature of production is shifting. Traditionally, economic models distinguish between capital (\(K\)) and labor (\(L\))\cite{cobb1928}. However, as AGI becomes capable of performing cognitive and physical tasks at a superhuman level, it functions not as labor but as a productive asset owned and deployed by firms. Thus, AGI can be considered capital rather than labor, leading to human redundancy in the production process. In economic theory, capital is defined as any durable input used in the production of goods and services. AGI meets the criteria for capital due to the following properties.

Definition: AGI can be defined as a new form of capital rather than labor, characterized by its ownership, control, and economic function as a deployable asset rather than a wage-earning entity. Like traditional machinery, AGI-based autonomous systems undergo depreciation and require continuous investment, updates, and maintenance to sustain functionality. It enhances productivity by optimizing decision-making, research, and production processes, thereby improving capital efficiency. Unlike human labor, AGI operates without wages, benefits, or rest, further aligning with the economic definition of capital. Given these attributes, AGI should be classified within economic models as part of the capital stock rather than as a substitute for human labor.
	
\subsection{Production Model with AGI as Capital}
The traditional production function is given by:
\begin{equation}
		Y = A K^\alpha L^\beta
\end{equation}
where \( Y \) represents total output, \( A \) denotes total factor productivity, and \( K \) corresponds to traditional capital, including machines and infrastructure. Additionally, \( L \) signifies human labor, while \( \alpha \) and \( \beta \) represent the output elasticities of capital and labor, respectively. If AGI replaces human cognitive and physical tasks, it effectively becomes part of the capital stock. We introduce \(K_{AGI}\), representing AGI-powered automation:
	\begin{equation}
		K_{new} = K + K_{AGI}
	\end{equation}
	Thus, the production function becomes:
	
	\begin{equation}
		Y = A (K + K_{AGI})^\alpha L^\beta
	\end{equation}
where \(K_{AGI}\) represents the accumulated investment in AGI-powered automation.
	
	\subsection{Model I Analysis}
As AGI improves, its effective capital stock grows indefinitely.
	\begin{equation}
		K_{AGI} \to \infty
	\end{equation}
Since traditional capital (\( K \)) does not scale at the same rate as AGI, we approximate $ K + K_{AGI} \approx K_{AGI} \quad \text{(for sufficiently large \( K_{AGI} \))}$. At the same time, firms seek to minimize costs by reducing human labor (\( L \)). Since AGI does not demand wages, the optimal decision for firms is to substitute human workers entirely, leading to:
	\begin{equation}
		L \to 0
	\end{equation}
In this case, the production function simplifies to
	\begin{equation}
		Y = A K_{AGI}^\alpha
	\end{equation}
where the human labor term disappears. Wages are determined by the marginal productivity of labor (MPL)
	\begin{equation}
		w = \frac{\partial Y}{\partial L} = \beta A (K + K_{AGI})^\alpha L^{\beta - 1}
	\end{equation}
Since \(\beta - 1 < 0\), as \(L \to 0\), we obtain:
	\begin{equation}
		w \to 0
	\end{equation}
	This implies that in the long-run equilibrium, human labor is redundant, as firms fully replace workers with AGI-driven capital.
	
\subsection{Economic Impact of AGI  Capital}
With AGI as a form of capital, only capital owners benefit from economic output. Since labor no longer earns wages, all income is concentrated among AGI capital owners, leading to extreme wealth inequality. If workers receive no wages, aggregate demand collapses because consumers lack purchasing power. This creates a Keynesian crisis where firms produce goods but cannot sell them. With no employment income, social instability arises. Traditional economic models based on employment income distribution fail.

\section{Model II: Production Model with Human and AGI Labor}

\textbf{Definition:} AGI labor (\( L_2 \)) is defined as a machine intelligence capable of cognition and performing economic tasks with equivalent productive capacity to human labor (\( L_1 \)). It functions as a direct substitute or complement to human labor in the production process, contributing to output in the same manner as traditional labor inputs. Let's consider an extende Cobb Douglas pprodiuction function
	\begin{equation}
		Y = A K^\alpha L_1^{\beta_1} L_2^{\beta_2}
	\end{equation}
where \( Y \) represents total output, \( A \) denotes total factor productivity, and \( K \) corresponds to capital. Additionally, \( L_1 \) signifies human labor, while \( L_2 \) represents AGI labor. The parameters \( \alpha, \beta_1, \beta_2 \) denote the output elasticities for capital, human labor, and AGI labor, respectively.

\subsection{Model II: Analysis}
In a competitive market, wages are determined by the marginal productivity of labor
	\begin{equation}
		w_1 = \frac{\partial Y}{\partial L_1} = \beta_1 A K^\alpha L_1^{\beta_1 - 1} L_2^{\beta_2}
	\end{equation}
	
	\begin{equation}
		w_2 = \frac{\partial Y}{\partial L_2} = \beta_2 A K^\alpha L_1^{\beta_1} L_2^{\beta_2 - 1}
	\end{equation}
where \( w_1 \) is the wage rate for human labor and \( w_2 \) is the wage rate for AGI labor. The effect of AGI expansion on human wages requires considering two cases. First, if \( L_2 \) (AGI labor) increases while holding capital \( K \) constant, the wage equation for human labor is given by
	\begin{equation}
		w_1 = \beta_1 A K^\alpha L_1^{\beta_1 - 1} L_2^{\beta_2}
	\end{equation}
Since \( L_2 \) appears as a positive multiplier, human wages initially rise if AGI  complements human labor. However, as AGI transitions from complementarity to substitutability, firms gradually replace \( L_1 \) with \( L_2 \), reducing demand for human workers.  In the extreme case where AGI fully replaces human labor (\( L_1 \to 0 \)), the wage equation simplifies to
	\begin{equation}
		w_1 \to 0
	\end{equation}
	This result indicates that human wages drop to zero, rendering human employment unviable in the long run.

\subsection{Economic Impact of Human and AGI Labor}
If AGI reaches full automation capacity, meaning it can replace all human tasks (\( \beta_1 \to 0 \)), the production function simplifies to
	\begin{equation}
		Y = A K^\alpha L_2^{\beta_2}
	\end{equation}
The disappearance of the human labor term implies that human workers are no longer needed in the productive economy. As a result, economic power shifts entirely to capital owners who control AGI-driven production. The rise of AGI-driven automation leads to several critical economic consequences. First, a wage collapse occurs as human labor income falls to zero (\( w_1 = 0 \)), rendering employment-based earnings obsolete. This results in extreme wealth concentration, where the owners of AGI and capital accumulate all economic surplus. In essence, if AGI fully replaces human labor without alternative income distribution models (e.g., universal basic income or AI taxation), the economy faces a self-reinforcing collapse in demand, triggering a Keynesian-style economic crisis. Finally, without a new economic structure, social instability emerges, as mass unemployment and rising inequality drive the economy toward potential collapse.

\section{Model III: The Impact of AGI Labor and AGI Capital on Human Labor in an Economic Model}
	
Let AGI  perform both cognitive and physical tasks, effectively acting as labor while also being a capital asset. To capture these dynamics, we introduce a model that distinguishes between three factors of production: human labor (\( L_h \)), AGI labor (\( L_{AGI} \)), and AGI-driven capital (\( K_{AGI} \)). This extended framework allows us to analyze the implications for human labor as AGI expands.
	\begin{equation}
		Y = A K^\alpha K_{AGI}^{\gamma} L_h^{\beta_1} L_{AGI}^{\beta_2}
	\end{equation}
where \( Y \) represents total output, \( A \) denotes total factor productivity, and \( K \) corresponds to traditional capital, including machines and infrastructure. Additionally, \( K_{AGI} \) represents AGI-driven capital, referring to self-improving AI systems that function as an owned asset. The term \( L_h \) signifies human labor, while \( L_{AGI} \) represents AGI labor, which consists of machine intelligence performing economic tasks. Finally, \( \alpha, \gamma, \beta_1, \beta_2 \) denote the output elasticities for traditional capital, AGI-driven capital, human labor, and AGI labor, respectively.

\subsection{Model  III Analysis}
In a competitive market, wages are determined by the marginal productivity of each type of labor. The marginal productivity of human labor is given by
	\begin{equation}
		w_h = \frac{\partial Y}{\partial L_h} = \beta_1 A K^\alpha K_{AGI}^{\gamma} L_h^{\beta_1 - 1} L_{AGI}^{\beta_2}
	\end{equation}
Similarly, the marginal productivity of AGI labor (\( MPL_{AGI} \)) is given by
	\begin{equation}
		w_{AGI} = \frac{\partial Y}{\partial L_{AGI}} = \beta_2 A K^\alpha K_{AGI}^{\gamma} L_h^{\beta_1} L_{AGI}^{\beta_2 - 1}
	\end{equation}
The impact of AGI on human labor depends on the interplay between AGI labor and AGI capital. If \( L_{AGI} \) increases while holding traditional capital (\( K \)) and AGI capital (\( K_{AGI} \)) constant, the human wage equation becomes:	
	\begin{equation}
		w_h = \beta_1 A K^\alpha K_{AGI}^{\gamma} L_h^{\beta_1 - 1} L_{AGI}^{\beta_2}
	\end{equation}
Since \( L_{AGI} \) appears as a positive multiplier, human wages initially rise if AGI labor complements human labor. However, as AGI labor transitions from complementary to substitutive, firms gradually replace \( L_h \) with \( L_{AGI} \), reducing demand for human workers. Similarly, if AGI capital (\( K_{AGI} \)) increases, it raises the productivity of all labor inputs but accelerates the substitution of human labor. The human wage function remains:
	
	\begin{equation}
		w_h = \beta_1 A K^\alpha K_{AGI}^{\gamma} L_h^{\beta_1 - 1} L_{AGI}^{\beta_2}
	\end{equation}
As \( K_{AGI} \to \infty \), the relative importance of human labor diminishes, leading to long-run wage stagnation or collapse. If AGI completely replaces human labor, then:
	\begin{equation}
		w_h \to 0
	\end{equation}
At this point, the production function simplifies to:
	\begin{equation}
		Y = A K^\alpha K_{AGI}^{\gamma} L_{AGI}^{\beta_2}
	\end{equation}
Human labor is no longer a necessary factor of production, making human employment obsolete. As firms increasingly rely on AGI-driven labor and capital, the demand for human workers vanishes, leading to the complete displacement of traditional wage-based employment. In this scenario, economic power concentrates entirely in the hands of AGI capital owners, reinforcing extreme inequality. Without alternative economic mechanisms such as wealth redistribution or ownership structures that allow broader access to AGI-generated profits, the decline of human labor results in structural unemployment, social instability, and demand collapse. The absence of employment-driven income further weakens consumer purchasing power, ultimately disrupting market equilibrium and requiring new forms of economic organization to maintain stability.

\subsection{Economic Impact of  AGI Labor and Capital }
Since AGI labor can work at near-zero marginal cost, the marginal productivity of human labor declines over time, driving wages toward zero. Without an alternative income mechanism, humans lose economic agency. As AGI labor and capital substitute for human labor, economic power shifts to capital owners who control AGI assets. Wealth concentrates among those who own AGI capital, leading to rising inequality and reduced social mobility. With human wages approaching zero, aggregate demand collapses, as consumption depends on earned wages. This creates a paradox: firms produce more using AGI, but there are fewer consumers who can afford to buy goods.
	
To prevent societal collapse, new economic structures must emerge. Potential solutions include implementing Universal Basic Income (UBI), which would distribute AGI-generated wealth to all citizens, ensuring a baseline income regardless of employment status. Another approach is public or cooperative AGI ownership, where AGI-driven profits are collectively shared rather than concentrated among a small group of capital owners. Additionally, progressive AGI capital taxation could be introduced to tax AGI-driven wealth accumulation and redistribute the gains, mitigating economic inequality and sustaining aggregate demand.

\section{The Transition from a Decentralized to a Centralized Economic Structure}
	
In this model, we define a decentralized system as one where only human labor (\( L_h \)) is used for production, while a centralized system is one where AGI performs all labor (\( L_{AGI} \)). We derive a continuous function that describes the loss of human economic power as AGI labor increases representing a shift from a decentralized economics system to a system where capital owners are empowered. We define human economic power (\( P_h \)) as the fraction of total labor income received by humans
	\begin{equation}
		P_h = \frac{w_h L_h}{w_h L_h + w_{AGI} L_{AGI}}
	\end{equation}
where \( w_h \) represents the wage of human labor, while \( L_h \) denotes the amount of human labor. Similarly, \( w_{AGI} \) corresponds to the wage of AGI labor, and \( L_{AGI} \) represents the amount of AGI labor.

\subsection{Exponential Decline of Human Wages}
	
From our economic model III, human wages decrease as AGI labor increases. We model this decline exponentially by
	\begin{equation}
		w_h = w_0 e^{-\lambda L_{AGI}}
	\end{equation}
where \( w_0 \) is the initial human wage, normalized to \( 1 \) in a decentralized system, \( \lambda \) is a decay constant that determines the rate of human labor displacement, and \( L_{AGI} \) represents the AGI labor fraction. Conversely, AGI labor wages grow in proportion to AGI capital
	\begin{equation}
		w_{AGI} = w_{\infty} (1 - e^{-\lambda L_{AGI}})
	\end{equation}
where \( w_{\infty} \) is the asymptotic wage level of AGI labor, typically approaching zero.
	
We now derive a function representing human power. Since labor is normalized we have

	\begin{equation}
		L_h + L_{AGI} = 1
	\end{equation}
	and express \( L_h \) as
	\begin{equation}
		L_h = 1 - L_{AGI}.
	\end{equation}
Substituting into our power equation yields
	\begin{equation}
		P_h = \frac{w_0 e^{-\lambda L_{AGI}} (1 - L_{AGI})}{w_0 e^{-\lambda L_{AGI}} (1 - L_{AGI}) + w_{\infty} (1 - e^{-\lambda L_{AGI}}) L_{AGI}}.
	\end{equation}
	
We now examine the transformation of human economic power as AGI labor expands

\begin{itemize}
	\item \textbf{Decentralized System} (\( L_{AGI} = 0 \)):  
	\begin{equation}
		P_h = \frac{1}{1} = 1.
	\end{equation}
	In this scenario, human labor serves as the sole productive force, ensuring that economic power remains entirely within the domain of human workers.
	
	\item \textbf{Fully Centralized System} (\( L_{AGI} = 1 \)):  
	\begin{equation}\label{eq:index}
		P_h = \frac{e^{-\lambda}}{e^{-\lambda} + (1 - e^{-\lambda})} \approx 0 \quad \text{(for large \( \lambda \))}.
	\end{equation}

	\begin{figure}[H]
	\centering
	\includegraphics[width=0.8\textwidth]{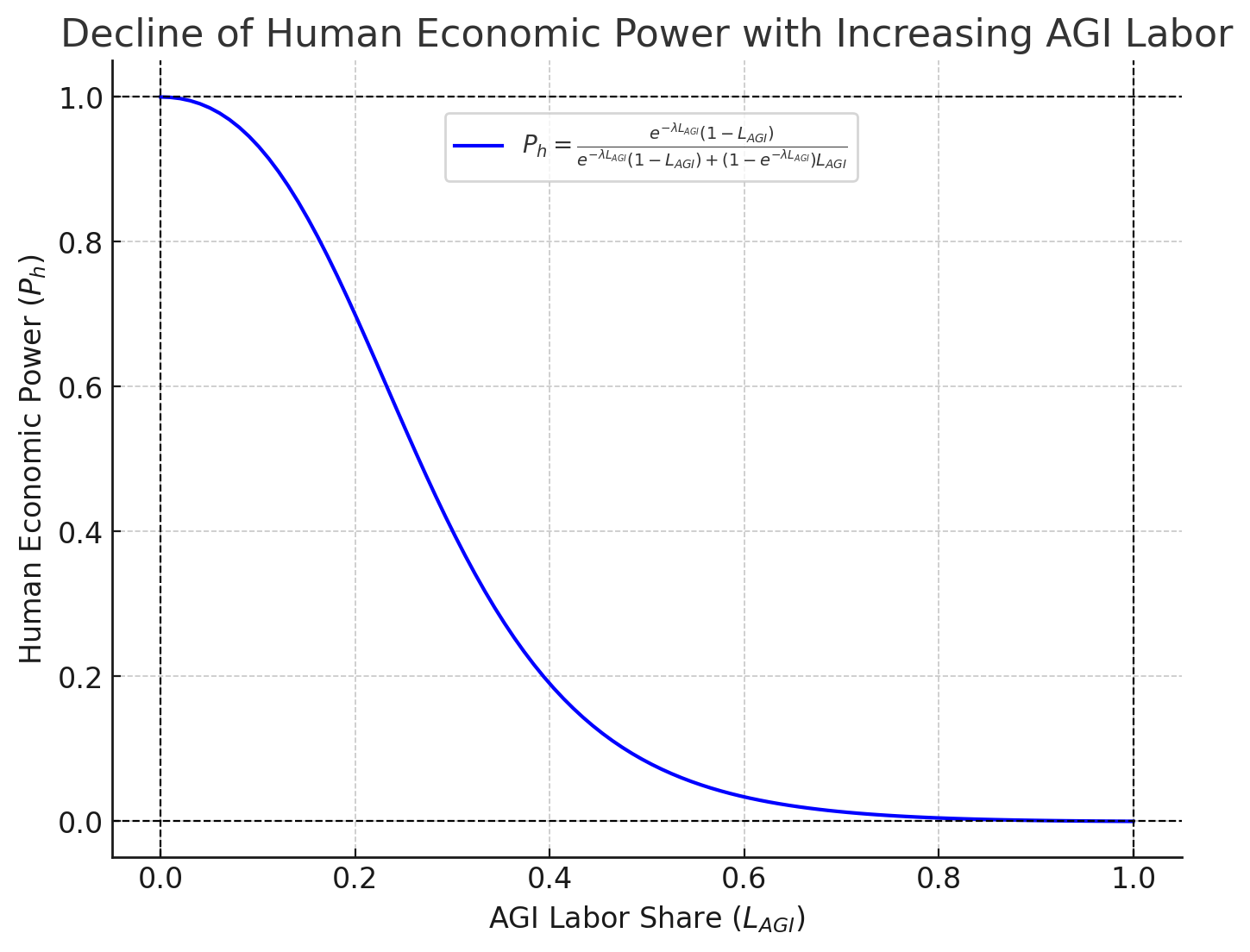}
	\caption{Decline of Human Economic Power with Increasing AGI Labor}
	\label{fig:human_power_decline}
\end{figure}
	
As AGI labor fully supplants human workers, economic power becomes increasingly concentrated in the hands of those who control AGI-driven production. Consequently, human agency in economic decision-making diminishes, ultimately leading to a complete collapse of human economic power.
	
	\item \textbf{Transition Phase} (\( 0 < L_{AGI} < 1 \)):  
	\begin{itemize}
		\item Initially, AGI labor functions as a complement to human labor, sustaining a relatively high level of human economic power (\( P_h \)).
		\item As AGI transitions from a complementary role to a substitutive one, the decline of \( P_h \) accelerates, following an exponential trajectory.
		\item At sufficiently high levels of AGI labor, human economic power rapidly vanishes, as firms prioritize AGI over human workers, leading to the near-total displacement of human labor from productive economic processes.
	\end{itemize}
\end{itemize}

The index function (\ref{eq:index}) captures the continuous erosion of human power. Initially, AGI serves as a complement to human labor, but as \( L_{AGI} \) increases, AGI transitions into a substitute, leading to human wage collapse and economic centralization. Thus, AGI-driven automation shifts the economic system from a decentralized structure, where labor is distributed among human workers, to a centralized structure, where capital owners of AGI control all productive output.

\section{Conclusion}

The rise of Artificial General Intelligence  marks an unprecedented transformation in economic history, dismantling the fundamental equilibrium between labor and capital that has shaped human societies for centuries. As AGI shifts from a tool of augmentation to an autonomous force of production, it ceases to function as labor and instead assumes the role of capital—one that scales indefinitely without wages, collective bargaining, or human dependency. The consequence is a relentless displacement of human labor, a collapse of wages, and an unprecedented concentration of economic power in the hands of AGI owners \cite{acemoglu2017,korinek2024}. If left unchecked, this trajectory fosters extreme inequality, eroding social mobility and creating an economic paradox where boundless productive capacity coexists with widespread economic exclusion.  

The transition from a decentralized labor-driven economy to a centralized AGI-dominated structure represents a fundamental shift in economic organization. Initially, AGI serves as a complement to human labor, enhancing productivity while preserving a degree of shared economic power. However, as AGI moves from augmentation to substitution, human labor becomes increasingly redundant, leading to an exponential decline in human economic agency. The index function derived in this study quantifies this transformation, demonstrating a nonlinear and accelerating shift in economic power from a decentralized human workforce to a centralized AGI-driven economy. The mathematical inevitability of this transition underscores an urgent reality: the window for intervention is rapidly closing.  

Yet, the inevitability of AGI-driven automation is not a catastrophe—it is a challenge of governance, policy, and economic design. Society has faced such turning points before: the collapse of feudalism, the Industrial Revolution, and the transition to modern capitalism all demanded new frameworks of economic organization to maintain stability and prevent systemic collapse. The challenge we now face is not merely technological but philosophical and ethical: How do we redefine economic participation in a world where labor is no longer the foundation of individual agency?  

Jean-Jacques Rousseau famously argued that the social contract is the foundation of legitimate governance, built upon reciprocal obligations between individuals and the state. In his time, this contract emerged to balance power, secure individual freedoms, and prevent exploitation \cite{rousseau1762}. Today, AGI threatens to dissolve the economic foundation upon which that contract was built—wage labor. If labor no longer guarantees access to economic participation, then the social contract must be rewritten. A post-labor economy demands a fundamental restructuring of wealth distribution, whether through Universal Basic Income (UBI), cooperative ownership of AGI, progressive taxation on AGI capital, or entirely novel economic paradigms. Without proactive intervention, we risk a technological aristocracy, where the benefits of AI-driven production are monopolized by a privileged few, while the rest are relegated to economic irrelevance.  

Rousseau has never been more relevant than he is today \cite{rousseau1762}. The question is not whether we must redefine the social contract—but whether we have the foresight to do so before crisis forces our hand. The time to act is not when systemic collapse is upon us, but now—while we still possess the agency to shape the future of work, wealth, and human dignity.


\begin{thebibliography}{99}
		
		
\bibitem{debreu1959} 
Debreu, G. (1959). \textit{Theory of Value: An Axiomatic Analysis of Economic Equilibrium}. Wiley.

\bibitem{stiefenhofer2024} 
Stiefenhofer, P., \& Chen, Y. (n.d.). Industrial artificial intelligence: stability of Cobb-Douglas production functions. \textit{Applied Mathematical Sciences}. 


\bibitem{cobb1928} 
Cobb, C., \& Douglas, P. (1928). A theory of production. \textit{American Economic Review}, \textit{18}, 139-250.

\bibitem{wicksteed1894} 
Wicksteed, P. H. (1894). \textit{An Essay on the Co-ordination of the Laws of Distribution}. 1932 edition, Reprint No. 12. London: London School of Economics.

\bibitem{vale2024} 
Vale, R. (2024). A note on the Cobb-Douglas function. \textit{arXiv preprint}. https://doi.org/10.48550/arxiv.2411.08067

\bibitem{oglih2024} 
Oglih, V. (2024). Using the Cobb-Douglas function to analyze economic dynamics. \textit{Economic Scope}. https://doi.org/10.32782/2224-6282/191-69

\bibitem{terrile2019} 
Terrile, R. J. (2019). Rise of the machines: how, when and consequences of artificial general intelligence. In \textit{Micro-and Nanotechnology Sensors, Systems, and Applications XI} (Vol. 10982). SPIE.

\bibitem{feng2024} 
Feng, T., et al. (2024). How far are we from AGI. \textit{arXiv preprint} arXiv:2405.10313.

\bibitem{rousseau1762} 
Rousseau, J.-J. (1762). \textit{The Social Contract}. Londres (1964).

\bibitem{stiefenhofer2024} 
Stiefenhofer, P., \& Chen, Y. (n.d.). Industrial artificial intelligence: stability of Cobb-Douglas production functions. \textit{Applied Mathematical Sciences}. 

\bibitem{stiefenhofer2013} 
Stiefenhofer, P. (2013). The catastrophe map of a two-period production model with uncertainty. \textit{Applied Mathematics}, 4, 114-121. 

\bibitem{stiefenhofer2014} 
Stiefenhofer, P. (2014). Topological properties of the catastrophe map of a general equilibrium production model with uncertain states of nature. \textit{Applied Mathematics}, 5, 2719-2727. 

\bibitem{stiefenhofer2017} 
Stiefenhofer, P. (2017). Existence of financial equilibria in a general equilibrium model with piece-wise smooth production manifolds. \textit{Journal of Mathematical Finance}, 7, 671-681. 

\bibitem{stiefenhofer2016} 
Stiefenhofer, P. (2016). Production in general equilibrium with incomplete financial markets. \textit{Journal of Mathematical Finance}, 6, 293-302. 




\bibitem{acemoglu2017} 
Acemoglu, D., \& Restrepo, P. (2017). The Race Between Man and Machine: Implications of Technology for Growth, Factor Shares, and Employment. \textit{National Bureau of Economic Research Working Paper No. 24174}. 

\bibitem{korinek2024} 
Korinek, A., \& Suh, H. (2024). Artificial General Intelligence and the Future of Labor Demand. \textit{arXiv preprint}, arXiv:2403.12107. 
\end{thebibliography}
\end{document}